\documentclass[aps,reprint,twocolumn,superscriptaddress,citeautoscript,showpacs,amsmath,amssymb,amsfonts,floatfix,prl]{revtex4-1}
\usepackage{gensymb}
\usepackage{graphicx}
\usepackage{dcolumn}
\usepackage{bm}
\usepackage{times}
\usepackage{color}

\begin{document}
\title{Helical magnetic ordering in Sr(Co$_{1-x}$Ni$_x$)$_2$As$_2$}

\author{J. M. Wilde}
\affiliation{Ames Laboratory, U. S. DOE, Ames, Iowa 50011, USA}
\affiliation{Department of Physics and Astronomy, Iowa State
University, Ames, Iowa 50011, USA}

\author{A. Kreyssig}
\affiliation{Ames Laboratory, U. S. DOE, Ames, Iowa 50011, USA}
\affiliation{Department of Physics and Astronomy, Iowa State
University, Ames, Iowa 50011, USA}

\author{D. Vaknin}
\affiliation{Ames Laboratory, U. S. DOE, Ames, Iowa 50011, USA}
\affiliation{Department of Physics and Astronomy, Iowa State
University, Ames, Iowa 50011, USA}

\author{N. S. Sangeetha}
\affiliation{Ames Laboratory, U. S. DOE, Ames, Iowa 50011, USA}
\affiliation{Department of Physics and Astronomy, Iowa State
University, Ames, Iowa 50011, USA}

\author{Bing Li}
\affiliation{Ames Laboratory, U. S. DOE, Ames, Iowa 50011, USA}
\affiliation{Department of Physics and Astronomy, Iowa State
University, Ames, Iowa 50011, USA}

\author{W.~Tian}
\affiliation{Neutron Scattering Division, Oak Ridge National Laboratory, Oak Ridge, TN 37831, USA}

\author{P.~P.~Orth}
\affiliation{Ames Laboratory, U. S. DOE, Ames, Iowa 50011, USA}
\affiliation{Department of Physics and Astronomy, Iowa State University, Ames, Iowa 50011, USA}

\author{D. C. Johnston}
\affiliation{Ames Laboratory, U. S. DOE, Ames, Iowa 50011, USA}
\affiliation{Department of Physics and Astronomy, Iowa State
University, Ames, Iowa 50011, USA}

\author{B. G. Ueland}
\affiliation{Ames Laboratory, U. S. DOE, Ames, Iowa 50011, USA}
\affiliation{Department of Physics and Astronomy, Iowa State
University, Ames, Iowa 50011, USA}

\author{R. J. McQueeney}
\affiliation{Ames Laboratory, U. S. DOE, Ames, Iowa 50011, USA}
\affiliation{Department of Physics and Astronomy, Iowa State
University, Ames, Iowa 50011, USA}

\date{\today}

\begin{abstract}
SrCo$_2$As$_2$ is a peculiar itinerant magnetic system that does not order magnetically, but inelastic neutron scattering experiments observe the same stripe-type antiferromagnetic (AF) fluctuations found in many of the Fe-based superconductors along with evidence of magnetic frustration. Here we present results from neutron diffraction measurements on single crystals of Sr(Co$_{1-x}$Ni$_x$)$_2$As$_2$ that show the development of long-range AF order with Ni-doping. However, the AF order is not stripe-type.  Rather, the magnetic structure consists of ferromagnetically-aligned (FM) layers (with moments laying in the layer) that are AF arranged along $\mathbf{c}$ with an incommensurate propagation vector of (0~0~$\tau$), i.e.\ a helix. Using high-energy x-ray diffraction, we find no evidence for a temperature-induced structural phase transition that would indicate a collinear AF order. This finding supports a picture of competing FM and AF interactions within the square transition-metal layers due to flat-band magnetic instabilities. However, the composition dependence of the propagation vector suggests that far more subtle Fermi surface and orbital effects control the interlayer magnetic correlations.
\end{abstract}


\maketitle
The Fe-based superconductors and their parent compounds \cite{Johnston_2010,Canfield_2010,Paglione_2010,Stewart_2011} are prime examples of intertwined structural, magnetic, and electronic ground states that can be sensitively tuned by chemical substitution \cite{Sefat_2008,Ni_2008, Li_2009,Canfield_2009, Ni_2009, Han_2009, Saha_2010}.  Phenomena emerging from these compounds such as spin and electronic nematic phases \cite{Fernandes_2012, Glasbrenner_2015}, magnetic frustration \cite{Si_2008, Yin_2011}, and magnetostructural volume-collapse transitions \cite{Kreyssig_2008, Goldman_2009, Ran_2011, Saha_2012}, and their interrelationships with superconductivity are central issues in condensed matter physics.  Such properties can often be interpreted in terms of either itinerant spin-density-wave (SDW) type or local-moment magnetism \cite{Xu_2008,Wyosocki_2011,Yamada_2013,Glasbrenner_2015} since the compounds are placed somewhere in between these two standard descriptions. Many of these phenomena extend in unique ways to the structurally related $A$Co$_2$As$_2$, $A=$~Ca, Sr, Ba, Eu, cobalt arsenides.  While the cobalt-arsenide materials are not found to be superconducting, they harbor signatures of weakly itinerant ferromagnetism (FM) \cite{Pandey_2013,Li_2019_Dai}, unusual spin fluctuations \cite{Jayasekara_2013, Sapkota_2017}, and magnetic frustration \cite{Sapkota_2017,Li_2019_CaSr}, which are tied to flat-band-driven Stoner instabilities \cite{Mao_2018, Li_2019_Dai}. 

Among the cobalt arsenides, tetragonal SrCo$_2$As$_2$ is unique.  While no long-range magnetic order is found, neutron scattering measurements find evidence for antiferromagnetic (AF) stripe-type spin fluctuations similar to those  associated with superconducting pairing in the Fe-based superconductors rather than the expected FM fluctuations \cite{Jayasekara_2013}.  In principle, moving the SrCo$_2$As$_2$ system closer to stripe-AF order through appropriate chemical substitution may realize superconductivity.  However, to date no cobalt-arsenide material has demonstrated long-range stripe-AF order  \cite{Jayasekara_2013,Jayasekara_2017, Li_2019_CaSr, Quirinale_2013}. 

\begin{figure}
	\centering\includegraphics[width=1\linewidth]{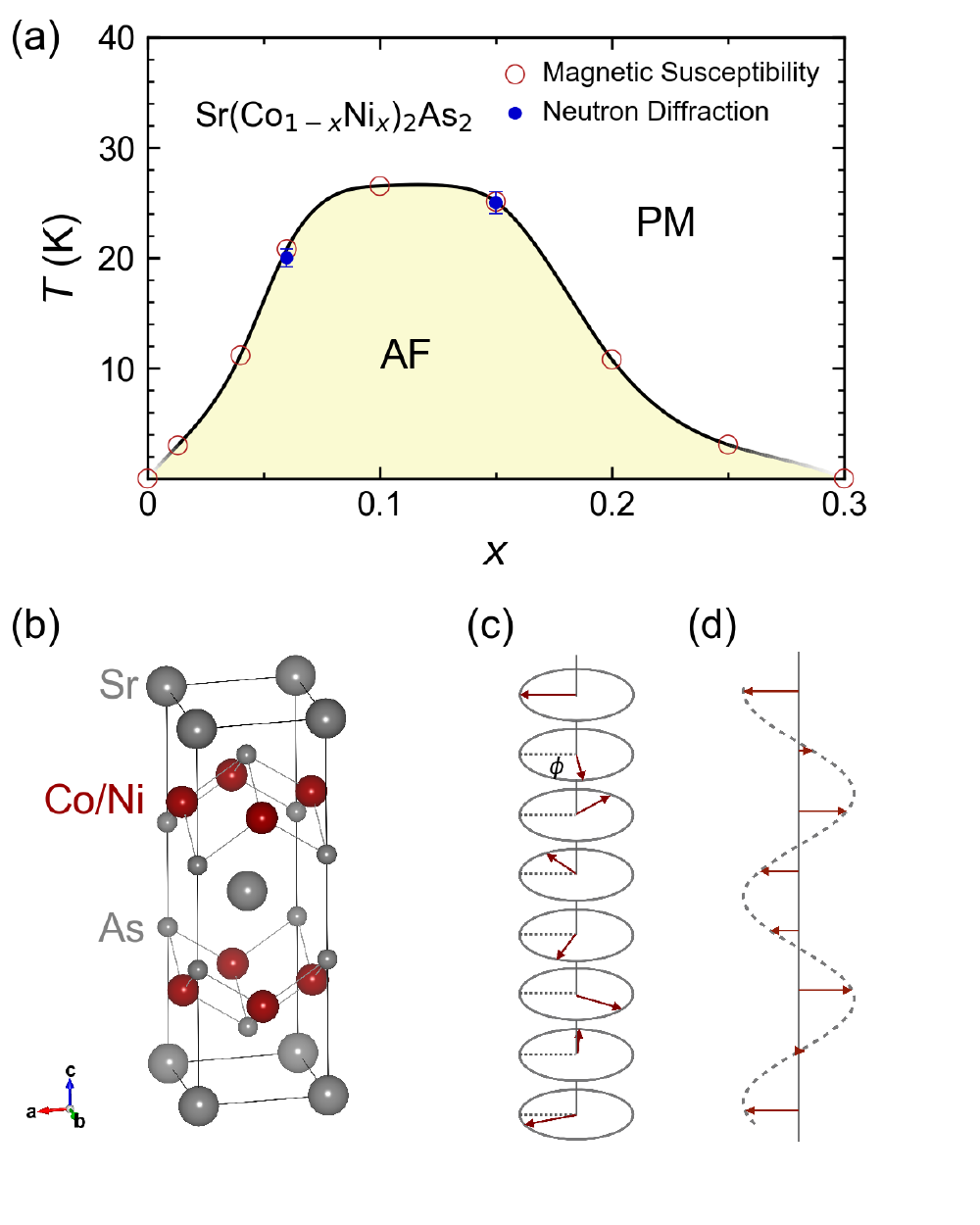}
	\caption{(a) Magnetic phase diagram for Sr(Co$_{1-x}$Ni$_x$)$_2$As$_2$ showing the antiferromagnetically ordered (AF) and paramagnetic (PM) phases. Solid (open) symbols correspond to neutron diffraction (magnetic susceptibility  \cite{Sangeetha_2019}) data.  (b) The chemical unit cell, in which magnetic moments exists for the Co/Ni sites. (c) Possible helical-AF arrangement of the magnetic moments with a turn angle of $\phi$ between ferromagnetically-aligned layers. (d) Possible collinear spin-density-wave type order, which cannot be differentiated from helical-AF order shown in (c) using neutron diffraction data.}
	\label{Fig1}
\end{figure}

Recently, the electron-doped materials Sr$_{1-x}$La$_x$Co$_2$As$_2$ \cite{Shen_2018} and Sr(Co$_{1-x}$Ni$_x$)$_2$As$_2$ \cite{Sangeetha_2019,Li_2019_SrCoNi} have been shown to develop long-range AF order for extremely low substitutions of $x \approx 1$\% as shown by the magnetic phase diagram in Fig.~\ref{Fig1}(a) for the Ni-doped series. A diagram of the chemical unit cell is given in Fig.~\ref{Fig1}(b).  Here, we study the Ni-doped compounds using neutron and high-energy x-ray diffraction to determine the microscopic details of the AF ground state.  Rather than the stripe-AF order one may expect from the spin-fluctuation spectrum of SrCo$_2$As$_2$ \cite{Jayasekara_2013,Li_2019_Dai,Li_2019_SrINS}, we find that the series develops \textit{incommensurate} AF order consisting of FM-aligned transition-metal layers where the in-plane ordered magnetic moment ($\bm{\mu}\perp\mathbf{c})$ forms a helix propagating perpendicular to the layers (i.e.\ along $\mathbf{c}$).  These results support recent evidence that SrCo$_2$As$_2$ possesses frustrated magnetic interactions driven by flat-band instabilities that place the system on the border between itinerant $2$D-FM and stripe-AF \cite{Li_2019_SrINS,Li_2019_Dai}.  Far more subtle, composition-dependent variations in interlayer interactions and magnetic anisotropy, as recently observed in Sr$_{1-x}$Ca$_{x}$Co$_2$As$_2$, are at play in determining the details of the layer stacking (such as helical, A-type, or more complex collinear magnetic order) \cite{Li_2019_CaSr}.    

Detailed sample growth and characterization data of Sr(Co$_{1-x}$Ni$_{x}$)$_{2}$As$_{2}$ have recently been described in Ref.~[\onlinecite{Sangeetha_2019}].  Whereas no AF order is found for $x=$ 0  down to $T=50$~mK \cite{Li_2019_SrINS}, magnetization and electronic-transport data indicate that small amounts of Ni-doping trigger AF order.  The dome of AF order spans $x \approx 0.013$--$0.25$ with a maximum N\'{e}el temperature of $T_{\text{N}} \approx27$~K.  We performed neutron and x-ray diffraction on $x= 0.06(1)$ and $0.15(2)$ single-crystal samples with $T_{\text{N}}= 20$ and $25$~K, respectively, as shown in Fig.~\ref{Fig1}(a).

Neutron diffraction measurements were performed on $\approx18$~mg single-crystal samples using the FIE-TAX diffractometer at the High Flux Isotope Reactor, Oak Ridge National Laboratory.  Samples were sealed in an Al can containing He exchange gas which was subsequently attached to the cold head of a closed-cycle He refrigerator. The beam collimators placed before the monochromator, between the the monochromator and sample, between the sample and analyzer, and between the analyzer and detector were $40^\prime$-$40^\prime$-$40^\prime$-$80^\prime$, respectively. FIE-TAX operates at a fixed incident energy of $14.7$~meV using two pyrolytic graphite (PG) monochromators. In order to significantly reduce higher harmonics in the incident beam, PG filters were mounted before and after the second monochromator. The scattering data are described using reciprocal lattice units of $\frac{2\pi}{a}$ for $H$ and $K$ and $\frac{2\pi}{c}$ for $L$ within the tetragonal ThCr$_{2}$Si$_{2}$-type structure (space group $I4/mmm$), where $a \approx 3.94$~\AA\ and $c\approx 11.61$~\AA. A detailed dependence of the lattice parameters on $x$ is given in Ref.~[\onlinecite{Sangeetha_2019}].  The samples were aligned with their ($H~H~L$) reciprocal-lattice planes coincident with the spectrometer's scattering plane.

High-energy x-ray diffraction measurements were performed on station $6$-ID-D at the Advanced Photon Source, Argonne National Laboratory. Measurements were made using $100$~keV x-rays, with the incident beam's direction normal to the $(H~H~L)$ and $(H~K~0)$ reciprocal-lattice planes. Diffraction patterns were recorded using a MAR$345$ area detector. Unlike lab sources, high-energy x-rays ensure that the bulk of the sample is probed.  By rocking the sample through small angular ranges about the axes perpendicular to the incident beam, we obtain an image of the reciprocal-lattice planes normal to the incident beam's direction \cite{Kreyssig_2007}.

We initially made neutron diffraction measurements at positions consistent with a stripe-AF propagation vector, $\textbf{q}_{\text{stripe}}= (\frac{1}{2}~\frac{1}{2}~L)$, and found that no magnetic Bragg peaks occur at $\textbf{q}_{\text{stripe}}$ for either $x=0.06$ and $0.15$. Rather, scans made along (0~0~$L$) and (1~1~$L$) revealed weak magnetic Bragg peaks at positions incommensurate with the chemical lattice.  Their positions are described by an AF propagation vector of (0~0~$\tau$), with $\tau=0.58(1)$ and $0.52(1)$ for $x=0.06$ and $0.15$, respectively. The highest intensity Bragg peaks occur at (0~0~$L\pm\tau$), $L$ even, and based on the sensitivity of neutrons to the component of $\bm{\mu}$ perpendicular to the scattering vector $\textbf{Q}$, we find that $\bm{\mu}$ lies within the $\mathbf{ab}$-plane for both compositions.  This result is consistent with magnetization data \cite{Sangeetha_2019}.  

\begin{figure*}
	\centering\includegraphics[width=1.00\linewidth]{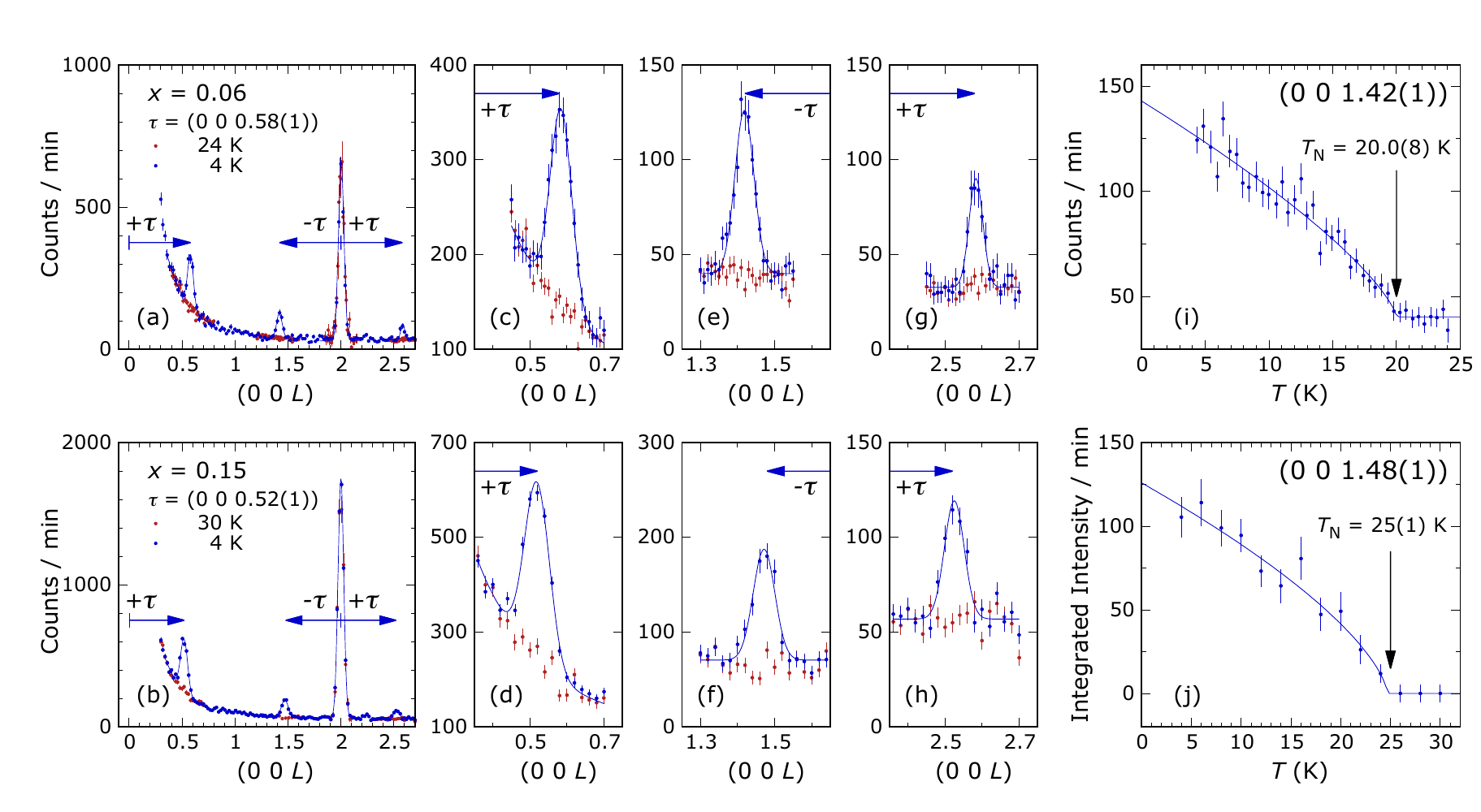}
	\caption{Neutron diffraction data for Sr(Co$_{1-x}$Ni$_x$)$_2$As$_2$ single crystals. (a) and (b) Scans along (0~0~$L$) for $x=0.06$ and $0.15$, respectively. (c)--(h) Expanded views of the magnetic Bragg peaks.  (i) and (j) Temperature dependence of the magnetic order parameter for $x=0.06$ and $0.15$, respectively, with lines as guides to the eye.}
	\label{Fig2}
\end{figure*}

Detailed scans along (0~0~$L$) at $T=4$~K are presented in Fig.~\ref{Fig2}, in which arrows point from the structural to the magnetic Bragg peaks. The widths of the magnetic Bragg peaks are resolution limited, which attests to the presence of long-range AF order, and scans along (1~1~$L$) yield similar results.  From measurements made at multiple temperatures, we find for both compositions that, within the resolution of our experiments, no significant change in $\tau$ or the widths of the magnetic Bragg peaks occur with temperature.  The temperature dependence of the integrated intensities of the (0~0~2-$\tau$) Bragg peaks for $x=0.06$ and $0.15$  are shown in Figs.~\ref{Fig2}(i) and \ref{Fig2}(j), and are consistent with a second-order AF transition with $T_{\text{N}}=20$ and $25$~K, respectively. These values agree with $T_{\text{N}}$ obtained from magnetization data \cite{Sangeetha_2019}, as shown in Fig.~\ref{Fig1}(a).  Overall, the result indicate that instead of the expected stripe-AF order, long-range incommensurate AF order develops in SrCo$_2$As$_2$ upon electron doping via partial replacement of Co with Ni.

Analysis of the diffraction data establishes that the magnetic structures for both values of $x$ consist of $2$D-FM layers that are AF arranged along $\mathbf{c}$ and controlled by $\tau$.  Interestingly, for both compositions, (0~0~$\tau$) is close to the commensurate value of (0~0~$\frac{1}{2}$) found for Sr$_{1-x}$Ca$_x$Co$_2$As$_2$, $0.5\alt x\alt0.8$  \cite{Li_2019_CaSr}.  In that case, the magnetic order is described by a doubling of the conventional body-centered unit cell along $\mathbf{c}$, containing four ferromagnetic Co layers that are stacked $\leftarrow\leftarrow\rightarrow\rightarrow$ along $\mathbf{c}$ or follow a $90\degree$-clock model.

In a similar vein, analysis of our neutron diffraction data for Sr(Co$_{1-x}$Ni$_x$)$_2$As$_2$ determines that the magnetic structure for $x=0.06$ and $0.15$ is either a collinear SDW, where the \textit{magnitude} of $\bm{\mu}$ varies sinusoidally along $\mathbf{c}$, or a non-collinear helix where the \textit{direction} of $\bm{\mu}$ varies along $\mathbf{c}$.  These AF structures are shown in Figs.~\ref{Fig1}(c) and \ref{Fig1}(d).  Although we cannot use the neutron diffraction data alone to distinguish between these two possible AF structures due to the presence of domains, combining our neutron diffraction results with a molecular-field analysis of the anisotropic magnetic susceptibility \cite{Sangeetha_2019} and our high-energy x-ray diffraction data favors the helical-AF structure.

As shown in Fig.~\ref{Fig1}(c), the AF stacking of the FM-aligned layers in the helical-AF structure can be parameterized by a turn angle $\phi$ between each layer. The values of $\tau$ determined from our neutron diffraction data yield $\phi=104\degree(2)$ and $94\degree(2)$, respectively, whereas the molecular-field analysis of the anisotropic susceptibility \cite{Sangeetha_2019} finds that $\phi=93\degree$ and $70\degree$ for $x=0.06$ and $0.15$, respectively.  Surprisingly, the trend in the dependence of the turn angle, $\phi$, on the Ni concentration, $x$, is correctly given by the local-moment model analysis of the anisotropic susceptibility \cite{Sangeetha_2019} of this itinerant magnetic system, and the values of $\phi$ are somewhat close.  We further find from the neutron diffraction data that $\mu=0.13(2)$ and $0.20(2)~\mu_{\text{B}}/(\text{Co+Ni})$ for  $x=0.06$ and $0.15$, respectively, for the helical-AF structure.  These values are in quite good agreement with the values for the saturation moment of $\mu_{\text{sat}}=0.117(1)$ and $0.165(1)~\mu_{\text{B}}/(\text{Co+Ni})$ determined for $x=0.06$ and $0.15$, respectively, from magnetization measurements via considerations using an itinerant FM model \cite{Sangeetha_2019}.  The existence of left- and right-handed helical domains would not affect the intensity of the neutron diffraction peaks.  

\begin{figure}
   \centering\includegraphics[width=1\linewidth]{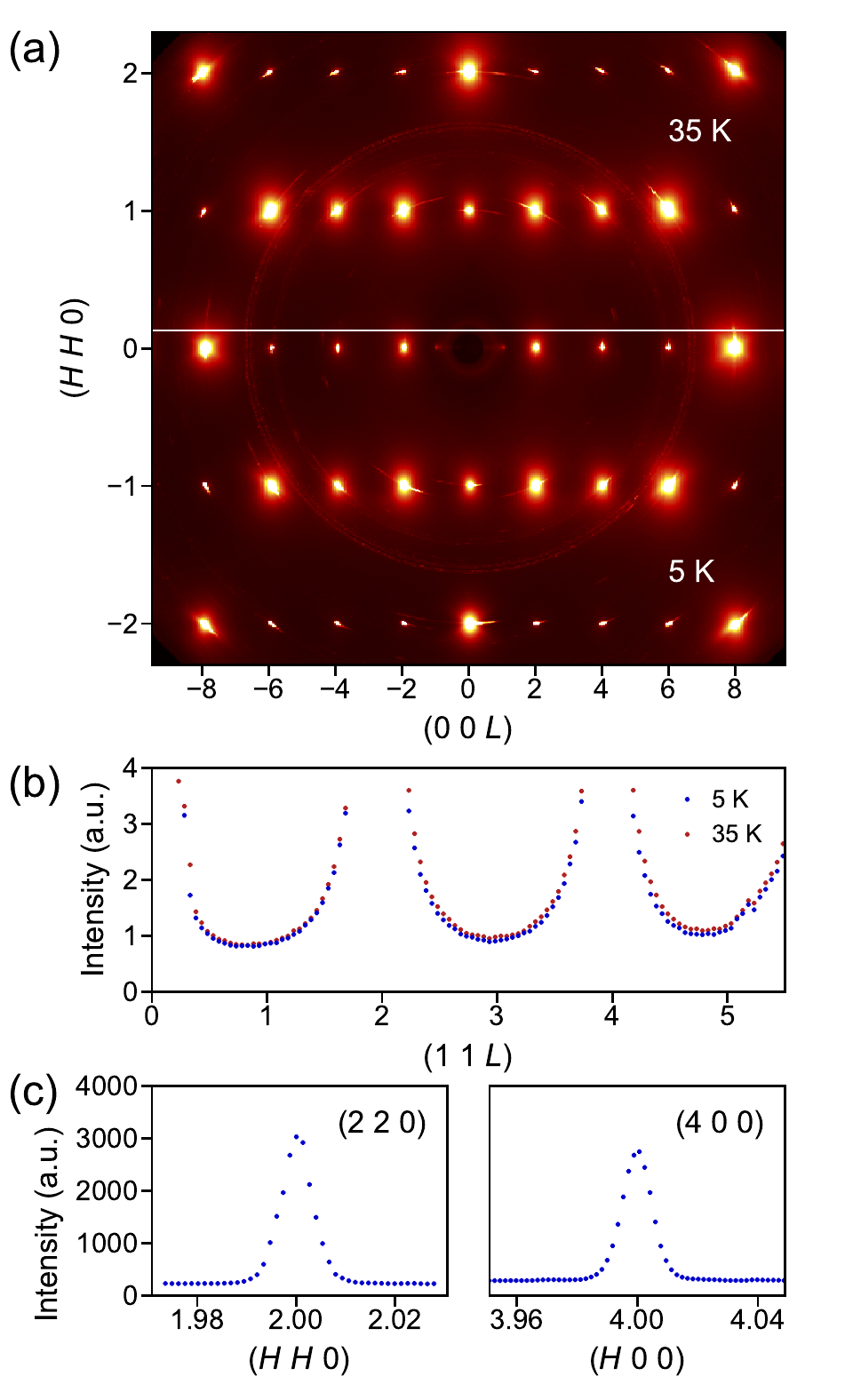}
   \caption{High-energy x-ray diffraction data for Sr(Co$_{1-x}$Ni$_{x}$)$_2$As$_2$ for $x= 0.05$.  (a) Image plots of data for the $(H~H~L)$ plane at $T = 35$ and $5$~K with intensities color coded. Two additional peaks close to the origin are due to the sample holder and are present at all temperatures.  (b) Cut along $(1~1~L)$ demonstrating the lack of CDW superstructure peaks.  (c) Cuts through the $(2~2~0)$ and $(4~0~0)$ Bragg peaks showing no splitting indicative of a potential orthorhombic lattice distortion.}
   \label{Fig3}
   \end{figure}

High-energy x-ray diffraction data were taken on a single-crystal sample of $x=0.05$ at $T=35$ and $5$~K to search for any structural anomalies associated with the AF ordering.  These data are shown in Fig.~\ref{Fig3}.  The $2$D images of the $(H~H~L)$ plane shown in Fig.~\ref{Fig3}(a) and the detailed cut along (1~1~$L$) in Fig.~\ref{Fig3}(b) show no additional Bragg peaks indicative of a superstructure or a charge-density wave, which is expected to accompany a SDW. In addition, no splitting of  Bragg peaks indicative of an orthorhombic lattice distortion are observed as shown in Fig.~\ref{Fig3}(c).  Such a distortion is typically expected for stripe-AF order or a  collinear SDW with $\bm{\mu}\perp\mathbf{c}$. Thus, analysis of these data favors the presence of helical-AF order.

Changes to the structure that accompany chemical substitutions are expected to affect (or reflect) the changing magnetic interactions.  In particular, CaCo$_{2}$As$_{2}$ has a value for $c$ that is $\approx13\%$ smaller than that for SrCo$_{2}$As$_{2}$ and exists in the collapsed-tetragonal (cT) phase \cite{Sangeetha_2017}.  CaCo$_{2}$As$_{2}$ shows A-type AF order with $\bm{\mu}\parallel\mathbf{c}$ \cite{Quirinale_2013}, the occurrence of which is tied to partially flat electronic bands lying closer to the Fermi energy $E_{\text{F}}$ than for paramagnetic SrCo$_{2}$As$_{2}$ \cite{Mao_2018}.  The value of $c$ changes with $x$ for both Sr(Co$_{1-x}$Ni$_{x}$)$_2$As$_2$ \cite{Sangeetha_2019,Li_2019_SrCoNi} and Sr$_{1-x}$Ca$_{x}$Co$_2$As$_2$ \cite{Sangeetha_2017}, and, interestingly, both series show regions of AF order with $\bm{\mu}\perp\mathbf{c}$ for ratios of $c/a$ corresponding to their uncollapsed-tetragonal (ucT) phases \cite{Sangeetha_2019,Sangeetha_2017}.  The ucT phase is characterized by weaker As-As covalent bonding than in the cT phase \cite{Hoffman_1985}.  However, the suppression of helical-AF order in Sr(Co$_{1-x}$Ni$_{x}$)$_2$As$_2$ with increasing $x$ \cite{Sangeetha_2019} and the emergence of AF order in Sr$_{1-x}$Ca$_{x}$Co$_2$As$_2$ do not appear to coincide with the ucT to cT crossover \cite{Sangeetha_2017}. Further, the mechanism controlling the direction of $\bm{\mu}$ is unclear.

Connections between the unique itinerant magnetic frustration and novel spin fluctuations in cobalt arsenides to partially flat electronic bands have recently been made using density-functional-theory calculations, angle-resolved photoemission spectroscopy, and elastic and inelastic neutron scattering experiments \cite{Mao_2018,Li_2019_SrINS,Li_2019_SrCoNi,Li_2019_CaSr, Sangeetha_2019}.  The flat bands originate from the transition metals' $3d_{x^2-y^2}$ orbitals \cite{Mao_2018,Sangeetha_2019} and form a sharp peak in the electronic density-of-states (DOS) close to $E_{\text{F}}$.   The proximity of the peak to $E_{\text{F}}$  is operative in determining the magnetism within the transition-metal planes via the Stoner mechanism \cite{Mao_2018}.  Thus, the capability to tune the \textit{intralayer} magnetism requires an understanding of the  effects of carrier doping, disorder (and the accompanying smearing of the DOS), and structural modulations (especially those affecting the degree of As-As covalent bonding \cite{Hoffman_1985}).   All of these effects can change the nature of the magnetism within the Co layers.  In this respect, the change from an intralayer stripe-AF instability in SrCo$_2$As$_2$ to a FM instability with small Ni substitution highlights the balance of several competing effects.

The nature by which the $2$D FM-aligned layers in AF ordered Sr(Co$_{1-x}$Ni$_{x}$)$_2$As$_2$  stack along $\mathbf{c}$ involves even more subtle \textit{interlayer} interactions and their susceptibility to chemical substitutions.  Similar complexity in the stacking of FM layers is reported for Sr$_{1-x}$Ca$_{x}$Co$_2$As$_2$ \cite{Li_2019_CaSr}.  In that case, the cascade of different AF stackings observed for high Ca concentrations is understood on the basis of a $1$D classical local-moment Heisenberg model known as the $J_0$-$J_1$-$J_2$ model \cite{Johnston_2012,Johnston_2015}, which is similar to the axial next-nearest-neighbor Ising (ANNNI) model \cite{Bak_1980, Fisher_1980,Villain_1980}.  For sufficiently large single-ion magnetic anisotropy, the $J_0$-$J_1$-$J_2$ model predicts FM ($\tau=0$), A-type AF ($\tau=1$,  $\uparrow\downarrow\uparrow\downarrow$), or phases with $\leftarrow\leftarrow\rightarrow\rightarrow$ or $90\degree$-clock AF structures ($\tau=\frac{1}{2}$).  Which phase exists depends on the ratio of the nearest-neighbor and next-nearest-neighbor interlayer interactions, $J_1$ and $J_2$, respectively.

For smaller values of magnetic anisotropy, these models admit incommensurate helical-AF phases, which are single-$\mathbf{Q}$ for the case of zero magnetic anisotropy.  Thus, in comparison to Sr$_{1-x}$Ca$_{x}$Co$_2$As$_2$, Sr(Co$_{1-x}$Ni$_{x}$)$_2$As$_2$ appears to fall into the regime of small magnetic anisotropy.  This is supported by the small value of the spin-flop fields \cite{Johnston_2015} seen for some of the AF ordered Ni-doped compounds \cite{Sangeetha_2019} in comparison to those for Sr$_{1-x}$Ca$_{x}$Co$_2$As$_2$ \cite{Sangeetha_2017,Li_2019_CaSr}.

Summarizing, our neutron diffraction results for Sr(Co$_{1-x}$Ni$_{x}$)$_2$As$_2$ show that its AF phase does not have stripe-AF order, but rather consists of FM-aligned transition-metal layers with $\bm{\mu}\perp\mathbf{c}$.  Unlike Sr$_{1-x}$Ca$_{x}$Co$_2$As$_2$, the AF arrangement along $c$ is incommensurate with the chemical lattice and characterized by (0~0~$\tau$), with $\tau=0.58(1)$ and $0.52(1)$ for $x=0.06$ and $0.15$, respectively.  The agreement between $\tau$ and a molecular-field analysis of anisotropic magnetic susceptibility data, as well as the absence of peaks corresponding to a charge-density wave and the lack of an orthorhombic lattice distortion in our high-energy x-ray diffraction data, tends to support a helical-AF structure rather than a collinear sinusoidally-modulated SDW, both of which are consistent with our neutron diffraction data.  Applying the helical-AF model to our neutron diffraction data yields $\mu=0.13(2)$ and $0.20(2)~\mu_{\text{B}}/(\text{Co+Ni})$ for  $x=0.06$ and $0.15$, respectively. Our results highlight that, in addition to the highly-tunable intralayer FM driven by the proximity of flat electronic bands to a van Hove singularity, the details of the AF stacking of the FM-aligned layers involve subtle interlayer magnetic interactions which are highly susceptible to doping.

It has been previously emphasized that a local-moment model for the itinerant AF CaCo$_{2-y}$As$_2$ accurately predicts the observed nearly-perfect magnetic frustration inferred from inelastic neutron-scattering measurements  \cite{Sapkota_2017}. Here, we have shown that its correspondence holds more generally as previously anticipated \cite{Sapkota_2017}. The helical AF structure predicted using a local-moment model for the magnetic susceptibility measurements in Ref.~[\onlinecite{Sangeetha_2019}] has been confirmed here not only qualitatively but also semi-quantitatively using magnetic neutron diffraction measurements.

\begin{acknowledgments}
	We are grateful for assistance from D.~S.~Robinson with performing the x-ray experiments and thank A.~I.~Goldman for his assistance and critical review of the manuscript.  We also thank D.~H.~Ryan, L.~Ke, and Y.~Sizyuk for helpful conversations.  Work at the Ames Laboratory was supported by the U.~S.~Department of Energy (DOE), Basic Energy Sciences, Division of Materials Sciences \& Engineering, under Contract No.~DE-AC$02$-$07$CH$11358$. A portion of this research used resources at the High Flux Isotope Reactor, a U.~S.~DOE Office of Science User Facility operated by the Oak Ridge National Laboratory.  This research used resources of the Advanced Photon Source, a U.~S.~DOE Office of Science User Facility operated for the U.~S.~DOE Office of Science by Argonne National Laboratory under Contract No.~DE-AC$02$-$06$CH$11357$. P.~P.~O. acknowledges support from Iowa State University Startup Funds.
\end{acknowledgments}

\bibliographystyle{apsrev4-1.bst}
\bibliography{SrCo1-xNix2As2.bib}

\begin{thebibliography}{42}%
\makeatletter
\providecommand \@ifxundefined [1]{%
 \@ifx{#1\undefined}
}%
\providecommand \@ifnum [1]{%
 \ifnum #1\expandafter \@firstoftwo
 \else \expandafter \@secondoftwo
 \fi
}%
\providecommand \@ifx [1]{%
 \ifx #1\expandafter \@firstoftwo
 \else \expandafter \@secondoftwo
 \fi
}%
\providecommand \natexlab [1]{#1}%
\providecommand \enquote  [1]{``#1''}%
\providecommand \bibnamefont  [1]{#1}%
\providecommand \bibfnamefont [1]{#1}%
\providecommand \citenamefont [1]{#1}%
\providecommand \href@noop [0]{\@secondoftwo}%
\providecommand \href [0]{\begingroup \@sanitize@url \@href}%
\providecommand \@href[1]{\@@startlink{#1}\@@href}%
\providecommand \@@href[1]{\endgroup#1\@@endlink}%
\providecommand \@sanitize@url [0]{\catcode `\\12\catcode `\$12\catcode
  `\&12\catcode `\#12\catcode `\^12\catcode `\_12\catcode `\%12\relax}%
\providecommand \@@startlink[1]{}%
\providecommand \@@endlink[0]{}%
\providecommand \url  [0]{\begingroup\@sanitize@url \@url }%
\providecommand \@url [1]{\endgroup\@href {#1}{\urlprefix }}%
\providecommand \urlprefix  [0]{URL }%
\providecommand \Eprint [0]{\href }%
\providecommand \doibase [0]{http://dx.doi.org/}%
\providecommand \selectlanguage [0]{\@gobble}%
\providecommand \bibinfo  [0]{\@secondoftwo}%
\providecommand \bibfield  [0]{\@secondoftwo}%
\providecommand \translation [1]{[#1]}%
\providecommand \BibitemOpen [0]{}%
\providecommand \bibitemStop [0]{}%
\providecommand \bibitemNoStop [0]{.\EOS\space}%
\providecommand \EOS [0]{\spacefactor3000\relax}%
\providecommand \BibitemShut  [1]{\csname bibitem#1\endcsname}%
\let\auto@bib@innerbib\@empty
\bibitem [{\citenamefont {Johnston}(2010)}]{Johnston_2010}%
  \BibitemOpen
  \bibfield  {author} {\bibinfo {author} {\bibfnamefont {D.~C.}\ \bibnamefont
  {Johnston}},\ }\href@noop {} {\bibfield  {journal} {\bibinfo  {journal} {Adv.
  Phys.}\ }\textbf {\bibinfo {volume} {59}},\ \bibinfo {pages} {803} (\bibinfo
  {year} {2010})}\BibitemShut {NoStop}%
\bibitem [{\citenamefont {Canfield}\ and\ \citenamefont
  {Bud'ko}(2010)}]{Canfield_2010}%
  \BibitemOpen
  \bibfield  {author} {\bibinfo {author} {\bibfnamefont {P.~C.}\ \bibnamefont
  {Canfield}}\ and\ \bibinfo {author} {\bibfnamefont {S.~L.}\ \bibnamefont
  {Bud'ko}},\ }\href@noop {} {\bibfield  {journal} {\bibinfo  {journal} {Annu.
  Rev. Condens. Matter Phys.}\ }\textbf {\bibinfo {volume} {1}},\ \bibinfo
  {pages} {27} (\bibinfo {year} {2010})}\BibitemShut {NoStop}%
\bibitem [{\citenamefont {Paglione}\ and\ \citenamefont
  {Greene}(2010)}]{Paglione_2010}%
  \BibitemOpen
  \bibfield  {author} {\bibinfo {author} {\bibfnamefont {J.}~\bibnamefont
  {Paglione}}\ and\ \bibinfo {author} {\bibfnamefont {R.~L.}\ \bibnamefont
  {Greene}},\ }\href@noop {} {\bibfield  {journal} {\bibinfo  {journal} {Nat.
  Phys.}\ }\textbf {\bibinfo {volume} {6}},\ \bibinfo {pages} {645} (\bibinfo
  {year} {2010})}\BibitemShut {NoStop}%
\bibitem [{\citenamefont {Stewart}(2011)}]{Stewart_2011}%
  \BibitemOpen
  \bibfield  {author} {\bibinfo {author} {\bibfnamefont {G.~R.}\ \bibnamefont
  {Stewart}},\ }\href@noop {} {\bibfield  {journal} {\bibinfo  {journal} {Rev.
  Mod. Phys.}\ }\textbf {\bibinfo {volume} {83}},\ \bibinfo {pages} {1589}
  (\bibinfo {year} {2011})}\BibitemShut {NoStop}%
\bibitem [{\citenamefont {Sefat}\ \emph {et~al.}(2008)\citenamefont {Sefat},
  \citenamefont {Jin}, \citenamefont {McGuire}, \citenamefont {Sales},
  \citenamefont {Singh},\ and\ \citenamefont {Mandrus}}]{Sefat_2008}%
  \BibitemOpen
  \bibfield  {author} {\bibinfo {author} {\bibfnamefont {A.~S.}\ \bibnamefont
  {Sefat}}, \bibinfo {author} {\bibfnamefont {R.}~\bibnamefont {Jin}}, \bibinfo
  {author} {\bibfnamefont {M.~A.}\ \bibnamefont {McGuire}}, \bibinfo {author}
  {\bibfnamefont {B.~C.}\ \bibnamefont {Sales}}, \bibinfo {author}
  {\bibfnamefont {D.~J.}\ \bibnamefont {Singh}}, \ and\ \bibinfo {author}
  {\bibfnamefont {D.}~\bibnamefont {Mandrus}},\ }\href {\doibase
  10.1103/PhysRevLett.101.117004} {\bibfield  {journal} {\bibinfo  {journal}
  {Phys. Rev. Lett.}\ }\textbf {\bibinfo {volume} {101}},\ \bibinfo {pages}
  {117004} (\bibinfo {year} {2008})}\BibitemShut {NoStop}%
\bibitem [{\citenamefont {Ni}\ \emph {et~al.}(2008)\citenamefont {Ni},
  \citenamefont {Bud'ko}, \citenamefont {Kreyssig}, \citenamefont {Nandi},
  \citenamefont {Rustan}, \citenamefont {Goldman}, \citenamefont {Gupta},
  \citenamefont {Corbett}, \citenamefont {Kracher},\ and\ \citenamefont
  {Canfield}}]{Ni_2008}%
  \BibitemOpen
  \bibfield  {author} {\bibinfo {author} {\bibfnamefont {N.}~\bibnamefont
  {Ni}}, \bibinfo {author} {\bibfnamefont {S.~L.}\ \bibnamefont {Bud'ko}},
  \bibinfo {author} {\bibfnamefont {A.}~\bibnamefont {Kreyssig}}, \bibinfo
  {author} {\bibfnamefont {S.}~\bibnamefont {Nandi}}, \bibinfo {author}
  {\bibfnamefont {G.~E.}\ \bibnamefont {Rustan}}, \bibinfo {author}
  {\bibfnamefont {A.~I.}\ \bibnamefont {Goldman}}, \bibinfo {author}
  {\bibfnamefont {S.}~\bibnamefont {Gupta}}, \bibinfo {author} {\bibfnamefont
  {J.~D.}\ \bibnamefont {Corbett}}, \bibinfo {author} {\bibfnamefont
  {A.}~\bibnamefont {Kracher}}, \ and\ \bibinfo {author} {\bibfnamefont
  {P.~C.}\ \bibnamefont {Canfield}},\ }\href {\doibase
  10.1103/PhysRevB.78.014507} {\bibfield  {journal} {\bibinfo  {journal} {Phys.
  Rev. B}\ }\textbf {\bibinfo {volume} {78}},\ \bibinfo {pages} {014507}
  (\bibinfo {year} {2008})}\BibitemShut {NoStop}%
\bibitem [{\citenamefont {Li}\ \emph {et~al.}(2009)\citenamefont {Li},
  \citenamefont {Luo}, \citenamefont {Wang}, \citenamefont {Chen},
  \citenamefont {Ren}, \citenamefont {Tao}, \citenamefont {Li}, \citenamefont
  {Lin}, \citenamefont {He}, \citenamefont {Zhu}, \citenamefont {Cao},\ and\
  \citenamefont {Xu}}]{Li_2009}%
  \BibitemOpen
  \bibfield  {author} {\bibinfo {author} {\bibfnamefont {L.~J.}\ \bibnamefont
  {Li}}, \bibinfo {author} {\bibfnamefont {Y.~K.}\ \bibnamefont {Luo}},
  \bibinfo {author} {\bibfnamefont {Q.~B.}\ \bibnamefont {Wang}}, \bibinfo
  {author} {\bibfnamefont {H.}~\bibnamefont {Chen}}, \bibinfo {author}
  {\bibfnamefont {Z.}~\bibnamefont {Ren}}, \bibinfo {author} {\bibfnamefont
  {Q.}~\bibnamefont {Tao}}, \bibinfo {author} {\bibfnamefont {Y.~K.}\
  \bibnamefont {Li}}, \bibinfo {author} {\bibfnamefont {X.}~\bibnamefont
  {Lin}}, \bibinfo {author} {\bibfnamefont {M.}~\bibnamefont {He}}, \bibinfo
  {author} {\bibfnamefont {Z.~W.}\ \bibnamefont {Zhu}}, \bibinfo {author}
  {\bibfnamefont {G.~H.}\ \bibnamefont {Cao}}, \ and\ \bibinfo {author}
  {\bibfnamefont {Z.~A.}\ \bibnamefont {Xu}},\ }\href {\doibase
  10.1088/1367-2630/11/2/025008} {\bibfield  {journal} {\bibinfo  {journal}
  {New J. Phys.}\ }\textbf {\bibinfo {volume} {11}},\ \bibinfo {pages} {025008}
  (\bibinfo {year} {2009})}\BibitemShut {NoStop}%
\bibitem [{\citenamefont {Canfield}\ \emph {et~al.}(2009)\citenamefont
  {Canfield}, \citenamefont {Bud'ko}, \citenamefont {Ni}, \citenamefont {Yan},\
  and\ \citenamefont {Kracher}}]{Canfield_2009}%
  \BibitemOpen
  \bibfield  {author} {\bibinfo {author} {\bibfnamefont {P.~C.}\ \bibnamefont
  {Canfield}}, \bibinfo {author} {\bibfnamefont {S.~L.}\ \bibnamefont
  {Bud'ko}}, \bibinfo {author} {\bibfnamefont {N.}~\bibnamefont {Ni}}, \bibinfo
  {author} {\bibfnamefont {J.~Q.}\ \bibnamefont {Yan}}, \ and\ \bibinfo
  {author} {\bibfnamefont {A.}~\bibnamefont {Kracher}},\ }\href {\doibase
  10.1103/PhysRevB.80.060501} {\bibfield  {journal} {\bibinfo  {journal} {Phys.
  Rev. B}\ }\textbf {\bibinfo {volume} {80}},\ \bibinfo {pages} {060501(R)}
  (\bibinfo {year} {2009})}\BibitemShut {NoStop}%
\bibitem [{\citenamefont {Ni}\ \emph {et~al.}(2009)\citenamefont {Ni},
  \citenamefont {Thaler}, \citenamefont {Kracher}, \citenamefont {Yan},
  \citenamefont {Bud'ko},\ and\ \citenamefont {Canfield}}]{Ni_2009}%
  \BibitemOpen
  \bibfield  {author} {\bibinfo {author} {\bibfnamefont {N.}~\bibnamefont
  {Ni}}, \bibinfo {author} {\bibfnamefont {A.}~\bibnamefont {Thaler}}, \bibinfo
  {author} {\bibfnamefont {A.}~\bibnamefont {Kracher}}, \bibinfo {author}
  {\bibfnamefont {J.~Q.}\ \bibnamefont {Yan}}, \bibinfo {author} {\bibfnamefont
  {S.~L.}\ \bibnamefont {Bud'ko}}, \ and\ \bibinfo {author} {\bibfnamefont
  {P.~C.}\ \bibnamefont {Canfield}},\ }\href {\doibase
  10.1103/PhysRevB.80.024511} {\bibfield  {journal} {\bibinfo  {journal} {Phys.
  Rev. B}\ }\textbf {\bibinfo {volume} {80}},\ \bibinfo {pages} {024511}
  (\bibinfo {year} {2009})}\BibitemShut {NoStop}%
\bibitem [{\citenamefont {Han}\ \emph {et~al.}(2009)\citenamefont {Han},
  \citenamefont {Zhu}, \citenamefont {Cheng}, \citenamefont {Mu}, \citenamefont
  {Jia}, \citenamefont {Fang}, \citenamefont {Wang}, \citenamefont {Luo},
  \citenamefont {Zeng}, \citenamefont {Shen}, \citenamefont {Shan},
  \citenamefont {Ren},\ and\ \citenamefont {Wen}}]{Han_2009}%
  \BibitemOpen
  \bibfield  {author} {\bibinfo {author} {\bibfnamefont {F.}~\bibnamefont
  {Han}}, \bibinfo {author} {\bibfnamefont {X.}~\bibnamefont {Zhu}}, \bibinfo
  {author} {\bibfnamefont {P.}~\bibnamefont {Cheng}}, \bibinfo {author}
  {\bibfnamefont {G.}~\bibnamefont {Mu}}, \bibinfo {author} {\bibfnamefont
  {Y.}~\bibnamefont {Jia}}, \bibinfo {author} {\bibfnamefont {L.}~\bibnamefont
  {Fang}}, \bibinfo {author} {\bibfnamefont {Y.}~\bibnamefont {Wang}}, \bibinfo
  {author} {\bibfnamefont {H.}~\bibnamefont {Luo}}, \bibinfo {author}
  {\bibfnamefont {B.}~\bibnamefont {Zeng}}, \bibinfo {author} {\bibfnamefont
  {B.}~\bibnamefont {Shen}}, \bibinfo {author} {\bibfnamefont {L.}~\bibnamefont
  {Shan}}, \bibinfo {author} {\bibfnamefont {C.}~\bibnamefont {Ren}}, \ and\
  \bibinfo {author} {\bibfnamefont {H.-H.}\ \bibnamefont {Wen}},\ }\href
  {\doibase 10.1103/PhysRevB.80.024506} {\bibfield  {journal} {\bibinfo
  {journal} {Phys. Rev. B}\ }\textbf {\bibinfo {volume} {80}},\ \bibinfo
  {pages} {024506} (\bibinfo {year} {2009})}\BibitemShut {NoStop}%
\bibitem [{\citenamefont {Saha}\ \emph {et~al.}(2010)\citenamefont {Saha},
  \citenamefont {Drye}, \citenamefont {Kirshenbaum}, \citenamefont {Butch},
  \citenamefont {Zavalij},\ and\ \citenamefont {Paglione}}]{Saha_2010}%
  \BibitemOpen
  \bibfield  {author} {\bibinfo {author} {\bibfnamefont {S.~R.}\ \bibnamefont
  {Saha}}, \bibinfo {author} {\bibfnamefont {T.}~\bibnamefont {Drye}}, \bibinfo
  {author} {\bibfnamefont {K.}~\bibnamefont {Kirshenbaum}}, \bibinfo {author}
  {\bibfnamefont {N.~P.}\ \bibnamefont {Butch}}, \bibinfo {author}
  {\bibfnamefont {P.~Y.}\ \bibnamefont {Zavalij}}, \ and\ \bibinfo {author}
  {\bibfnamefont {J.}~\bibnamefont {Paglione}},\ }\href {\doibase
  10.1088/0953-8984/22/7/072204} {\bibfield  {journal} {\bibinfo  {journal} {J.
  Phys.: Condens. Matter}\ }\textbf {\bibinfo {volume} {22}},\ \bibinfo {pages}
  {072204} (\bibinfo {year} {2010})}\BibitemShut {NoStop}%
\bibitem [{\citenamefont {Fernandes}\ \emph {et~al.}(2012)\citenamefont
  {Fernandes}, \citenamefont {Chubukov}, \citenamefont {Knolle}, \citenamefont
  {Eremin},\ and\ \citenamefont {Schmalian}}]{Fernandes_2012}%
  \BibitemOpen
  \bibfield  {author} {\bibinfo {author} {\bibfnamefont {R.~M.}\ \bibnamefont
  {Fernandes}}, \bibinfo {author} {\bibfnamefont {A.~V.}\ \bibnamefont
  {Chubukov}}, \bibinfo {author} {\bibfnamefont {J.}~\bibnamefont {Knolle}},
  \bibinfo {author} {\bibfnamefont {I.}~\bibnamefont {Eremin}}, \ and\ \bibinfo
  {author} {\bibfnamefont {J.}~\bibnamefont {Schmalian}},\ }\href {\doibase
  10.1103/PhysRevB.85.024534} {\bibfield  {journal} {\bibinfo  {journal} {Phys.
  Rev. B}\ }\textbf {\bibinfo {volume} {85}},\ \bibinfo {pages} {024534}
  (\bibinfo {year} {2012})}\BibitemShut {NoStop}%
\bibitem [{\citenamefont {Glasbrenner}\ \emph {et~al.}(2015)\citenamefont
  {Glasbrenner}, \citenamefont {Mazin}, \citenamefont {Jeschke}, \citenamefont
  {Hirschfeld}, \citenamefont {Fernandes},\ and\ \citenamefont
  {Valent{\'\i}}}]{Glasbrenner_2015}%
  \BibitemOpen
  \bibfield  {author} {\bibinfo {author} {\bibfnamefont {J.~K.}\ \bibnamefont
  {Glasbrenner}}, \bibinfo {author} {\bibfnamefont {I.~I.}\ \bibnamefont
  {Mazin}}, \bibinfo {author} {\bibfnamefont {H.~O.}\ \bibnamefont {Jeschke}},
  \bibinfo {author} {\bibfnamefont {P.~J.}\ \bibnamefont {Hirschfeld}},
  \bibinfo {author} {\bibfnamefont {R.~M.}\ \bibnamefont {Fernandes}}, \ and\
  \bibinfo {author} {\bibfnamefont {R.}~\bibnamefont {Valent{\'\i}}},\ }\href
  {https://doi.org/10.1038/nphys3434} {\bibfield  {journal} {\bibinfo
  {journal} {Nat. Phys.}\ }\textbf {\bibinfo {volume} {11}},\ \bibinfo {pages}
  {953 EP } (\bibinfo {year} {2015})}\BibitemShut {NoStop}%
\bibitem [{\citenamefont {Si}\ and\ \citenamefont {Abrahams}(2008)}]{Si_2008}%
  \BibitemOpen
  \bibfield  {author} {\bibinfo {author} {\bibfnamefont {Q.}~\bibnamefont
  {Si}}\ and\ \bibinfo {author} {\bibfnamefont {E.}~\bibnamefont {Abrahams}},\
  }\href@noop {} {\bibfield  {journal} {\bibinfo  {journal} {Phys. Rev. Lett.}\
  }\textbf {\bibinfo {volume} {101}},\ \bibinfo {pages} {076401} (\bibinfo
  {year} {2008})}\BibitemShut {NoStop}%
\bibitem [{\citenamefont {Yin}\ \emph {et~al.}(2011)\citenamefont {Yin},
  \citenamefont {Haule},\ and\ \citenamefont {Kotliar}}]{Yin_2011}%
  \BibitemOpen
  \bibfield  {author} {\bibinfo {author} {\bibfnamefont {Z.~P.}\ \bibnamefont
  {Yin}}, \bibinfo {author} {\bibfnamefont {K.}~\bibnamefont {Haule}}, \ and\
  \bibinfo {author} {\bibfnamefont {G.}~\bibnamefont {Kotliar}},\ }\href@noop
  {} {\bibfield  {journal} {\bibinfo  {journal} {Nat. Mat.}\ }\textbf {\bibinfo
  {volume} {10}},\ \bibinfo {pages} {932} (\bibinfo {year} {2011})}\BibitemShut
  {NoStop}%
\bibitem [{\citenamefont {Kreyssig}\ \emph {et~al.}(2008)\citenamefont
  {Kreyssig}, \citenamefont {Green}, \citenamefont {Lee}, \citenamefont
  {Samolyuk}, \citenamefont {Zajdel}, \citenamefont {Lynn}, \citenamefont
  {Bud'ko}, \citenamefont {Torikachvili}, \citenamefont {Ni}, \citenamefont
  {Nandi}, \citenamefont {Le\~ao}, \citenamefont {Poulton}, \citenamefont
  {Argyriou}, \citenamefont {Harmon}, \citenamefont {McQueeney}, \citenamefont
  {Canfield},\ and\ \citenamefont {Goldman}}]{Kreyssig_2008}%
  \BibitemOpen
  \bibfield  {author} {\bibinfo {author} {\bibfnamefont {A.}~\bibnamefont
  {Kreyssig}}, \bibinfo {author} {\bibfnamefont {M.~A.}\ \bibnamefont {Green}},
  \bibinfo {author} {\bibfnamefont {Y.}~\bibnamefont {Lee}}, \bibinfo {author}
  {\bibfnamefont {G.~D.}\ \bibnamefont {Samolyuk}}, \bibinfo {author}
  {\bibfnamefont {P.}~\bibnamefont {Zajdel}}, \bibinfo {author} {\bibfnamefont
  {J.~W.}\ \bibnamefont {Lynn}}, \bibinfo {author} {\bibfnamefont {S.~L.}\
  \bibnamefont {Bud'ko}}, \bibinfo {author} {\bibfnamefont {M.~S.}\
  \bibnamefont {Torikachvili}}, \bibinfo {author} {\bibfnamefont
  {N.}~\bibnamefont {Ni}}, \bibinfo {author} {\bibfnamefont {S.}~\bibnamefont
  {Nandi}}, \bibinfo {author} {\bibfnamefont {J.~B.}\ \bibnamefont {Le\~ao}},
  \bibinfo {author} {\bibfnamefont {S.~J.}\ \bibnamefont {Poulton}}, \bibinfo
  {author} {\bibfnamefont {D.~N.}\ \bibnamefont {Argyriou}}, \bibinfo {author}
  {\bibfnamefont {B.~N.}\ \bibnamefont {Harmon}}, \bibinfo {author}
  {\bibfnamefont {R.~J.}\ \bibnamefont {McQueeney}}, \bibinfo {author}
  {\bibfnamefont {P.~C.}\ \bibnamefont {Canfield}}, \ and\ \bibinfo {author}
  {\bibfnamefont {A.~I.}\ \bibnamefont {Goldman}},\ }\href@noop {} {\bibfield
  {journal} {\bibinfo  {journal} {Phys. Rev. B}\ }\textbf {\bibinfo {volume}
  {78}},\ \bibinfo {pages} {184517} (\bibinfo {year} {2008})}\BibitemShut
  {NoStop}%
\bibitem [{\citenamefont {Goldman}\ \emph {et~al.}(2009)\citenamefont
  {Goldman}, \citenamefont {Kreyssig}, \citenamefont
  {Proke\ifmmode~\check{s}\else \v{s}\fi{}}, \citenamefont {Pratt},
  \citenamefont {Argyriou}, \citenamefont {Lynn}, \citenamefont {Nandi},
  \citenamefont {Kimber}, \citenamefont {Chen}, \citenamefont {Lee},
  \citenamefont {Samolyuk}, \citenamefont {Le\~ao}, \citenamefont {Poulton},
  \citenamefont {Bud'ko}, \citenamefont {Ni}, \citenamefont {Canfield},
  \citenamefont {Harmon},\ and\ \citenamefont {McQueeney}}]{Goldman_2009}%
  \BibitemOpen
  \bibfield  {author} {\bibinfo {author} {\bibfnamefont {A.~I.}\ \bibnamefont
  {Goldman}}, \bibinfo {author} {\bibfnamefont {A.}~\bibnamefont {Kreyssig}},
  \bibinfo {author} {\bibfnamefont {K.}~\bibnamefont
  {Proke\ifmmode~\check{s}\else \v{s}\fi{}}}, \bibinfo {author} {\bibfnamefont
  {D.~K.}\ \bibnamefont {Pratt}}, \bibinfo {author} {\bibfnamefont {D.~N.}\
  \bibnamefont {Argyriou}}, \bibinfo {author} {\bibfnamefont {J.~W.}\
  \bibnamefont {Lynn}}, \bibinfo {author} {\bibfnamefont {S.}~\bibnamefont
  {Nandi}}, \bibinfo {author} {\bibfnamefont {S.~A.~J.}\ \bibnamefont
  {Kimber}}, \bibinfo {author} {\bibfnamefont {Y.}~\bibnamefont {Chen}},
  \bibinfo {author} {\bibfnamefont {Y.~B.}\ \bibnamefont {Lee}}, \bibinfo
  {author} {\bibfnamefont {G.}~\bibnamefont {Samolyuk}}, \bibinfo {author}
  {\bibfnamefont {J.~B.}\ \bibnamefont {Le\~ao}}, \bibinfo {author}
  {\bibfnamefont {S.~J.}\ \bibnamefont {Poulton}}, \bibinfo {author}
  {\bibfnamefont {S.~L.}\ \bibnamefont {Bud'ko}}, \bibinfo {author}
  {\bibfnamefont {N.}~\bibnamefont {Ni}}, \bibinfo {author} {\bibfnamefont
  {P.~C.}\ \bibnamefont {Canfield}}, \bibinfo {author} {\bibfnamefont {B.~N.}\
  \bibnamefont {Harmon}}, \ and\ \bibinfo {author} {\bibfnamefont {R.~J.}\
  \bibnamefont {McQueeney}},\ }\href@noop {} {\bibfield  {journal} {\bibinfo
  {journal} {Phys. Rev. B}\ }\textbf {\bibinfo {volume} {79}},\ \bibinfo
  {pages} {024513} (\bibinfo {year} {2009})}\BibitemShut {NoStop}%
\bibitem [{\citenamefont {Ran}\ \emph {et~al.}(2011)\citenamefont {Ran},
  \citenamefont {Bud'ko}, \citenamefont {Pratt}, \citenamefont {Kreyssig},
  \citenamefont {Kim}, \citenamefont {Kramer}, \citenamefont {Ryan},
  \citenamefont {Rowan-Weetaluktuk}, \citenamefont {Furukawa}, \citenamefont
  {Roy}, \citenamefont {Goldman},\ and\ \citenamefont {Canfield}}]{Ran_2011}%
  \BibitemOpen
  \bibfield  {author} {\bibinfo {author} {\bibfnamefont {S.}~\bibnamefont
  {Ran}}, \bibinfo {author} {\bibfnamefont {S.~L.}\ \bibnamefont {Bud'ko}},
  \bibinfo {author} {\bibfnamefont {D.~K.}\ \bibnamefont {Pratt}}, \bibinfo
  {author} {\bibfnamefont {A.}~\bibnamefont {Kreyssig}}, \bibinfo {author}
  {\bibfnamefont {M.~G.}\ \bibnamefont {Kim}}, \bibinfo {author} {\bibfnamefont
  {M.~J.}\ \bibnamefont {Kramer}}, \bibinfo {author} {\bibfnamefont {D.~H.}\
  \bibnamefont {Ryan}}, \bibinfo {author} {\bibfnamefont {W.~N.}\ \bibnamefont
  {Rowan-Weetaluktuk}}, \bibinfo {author} {\bibfnamefont {Y.}~\bibnamefont
  {Furukawa}}, \bibinfo {author} {\bibfnamefont {B.}~\bibnamefont {Roy}},
  \bibinfo {author} {\bibfnamefont {A.~I.}\ \bibnamefont {Goldman}}, \ and\
  \bibinfo {author} {\bibfnamefont {P.~C.}\ \bibnamefont {Canfield}},\ }\href
  {\doibase 10.1103/PhysRevB.83.144517} {\bibfield  {journal} {\bibinfo
  {journal} {Phys. Rev. B}\ }\textbf {\bibinfo {volume} {83}},\ \bibinfo
  {pages} {144517} (\bibinfo {year} {2011})}\BibitemShut {NoStop}%
\bibitem [{\citenamefont {Saha}\ \emph {et~al.}(2012)\citenamefont {Saha},
  \citenamefont {Butch}, \citenamefont {Drye}, \citenamefont {Magill},
  \citenamefont {Ziemak}, \citenamefont {Kirshenbaum}, \citenamefont {Zavalij},
  \citenamefont {Lynn},\ and\ \citenamefont {Paglione}}]{Saha_2012}%
  \BibitemOpen
  \bibfield  {author} {\bibinfo {author} {\bibfnamefont {S.~R.}\ \bibnamefont
  {Saha}}, \bibinfo {author} {\bibfnamefont {N.~P.}\ \bibnamefont {Butch}},
  \bibinfo {author} {\bibfnamefont {T.}~\bibnamefont {Drye}}, \bibinfo {author}
  {\bibfnamefont {J.}~\bibnamefont {Magill}}, \bibinfo {author} {\bibfnamefont
  {S.}~\bibnamefont {Ziemak}}, \bibinfo {author} {\bibfnamefont
  {K.}~\bibnamefont {Kirshenbaum}}, \bibinfo {author} {\bibfnamefont {P.~Y.}\
  \bibnamefont {Zavalij}}, \bibinfo {author} {\bibfnamefont {J.~W.}\
  \bibnamefont {Lynn}}, \ and\ \bibinfo {author} {\bibfnamefont
  {J.}~\bibnamefont {Paglione}},\ }\href@noop {} {\bibfield  {journal}
  {\bibinfo  {journal} {Phys. Rev. B}\ }\textbf {\bibinfo {volume} {85}},\
  \bibinfo {pages} {024525} (\bibinfo {year} {2012})}\BibitemShut {NoStop}%
\bibitem [{\citenamefont {Xu}\ \emph {et~al.}(2008)\citenamefont {Xu},
  \citenamefont {M{\"u}ller},\ and\ \citenamefont {Sachdev}}]{Xu_2008}%
  \BibitemOpen
  \bibfield  {author} {\bibinfo {author} {\bibfnamefont {C.}~\bibnamefont
  {Xu}}, \bibinfo {author} {\bibfnamefont {M.}~\bibnamefont {M{\"u}ller}}, \
  and\ \bibinfo {author} {\bibfnamefont {S.}~\bibnamefont {Sachdev}},\
  }\href@noop {} {\bibfield  {journal} {\bibinfo  {journal} {Phys. Rev. B}\
  }\textbf {\bibinfo {volume} {78}},\ \bibinfo {pages} {020501(R)} (\bibinfo
  {year} {2008})}\BibitemShut {NoStop}%
\bibitem [{\citenamefont {Wysocki}\ \emph {et~al.}(2011)\citenamefont
  {Wysocki}, \citenamefont {Belashchenko},\ and\ \citenamefont
  {Antropov}}]{Wyosocki_2011}%
  \BibitemOpen
  \bibfield  {author} {\bibinfo {author} {\bibfnamefont {A.~L.}\ \bibnamefont
  {Wysocki}}, \bibinfo {author} {\bibfnamefont {K.~D.}\ \bibnamefont
  {Belashchenko}}, \ and\ \bibinfo {author} {\bibfnamefont {V.~P.}\
  \bibnamefont {Antropov}},\ }\href {https://doi.org/10.1038/nphys1933}
  {\bibfield  {journal} {\bibinfo  {journal} {Nat. Phys.}\ }\textbf {\bibinfo
  {volume} {7}},\ \bibinfo {pages} {485 EP } (\bibinfo {year}
  {2011})}\BibitemShut {NoStop}%
\bibitem [{\citenamefont {Yamada}\ \emph {et~al.}(2013)\citenamefont {Yamada},
  \citenamefont {Seki}, \citenamefont {Eder},\ and\ \citenamefont
  {Ohta}}]{Yamada_2013}%
  \BibitemOpen
  \bibfield  {author} {\bibinfo {author} {\bibfnamefont {A.}~\bibnamefont
  {Yamada}}, \bibinfo {author} {\bibfnamefont {K.}~\bibnamefont {Seki}},
  \bibinfo {author} {\bibfnamefont {R.}~\bibnamefont {Eder}}, \ and\ \bibinfo
  {author} {\bibfnamefont {Y.}~\bibnamefont {Ohta}},\ }\href@noop {} {\bibfield
   {journal} {\bibinfo  {journal} {Phys. Rev. B}\ }\textbf {\bibinfo {volume}
  {88}},\ \bibinfo {pages} {075114} (\bibinfo {year} {2013})}\BibitemShut
  {NoStop}%
\bibitem [{\citenamefont {Pandey}\ \emph {et~al.}(2013)\citenamefont {Pandey},
  \citenamefont {Quirinale}, \citenamefont {Jayasekara}, \citenamefont
  {Sapkota}, \citenamefont {Kim}, \citenamefont {Dhaka}, \citenamefont {Lee},
  \citenamefont {Heitmann}, \citenamefont {Stephens}, \citenamefont
  {Ogloblichev}, \citenamefont {Kreyssig}, \citenamefont {McQueeney},
  \citenamefont {Goldman}, \citenamefont {Kaminski}, \citenamefont {Harmon},
  \citenamefont {Furukawa},\ and\ \citenamefont {Johnston}}]{Pandey_2013}%
  \BibitemOpen
  \bibfield  {author} {\bibinfo {author} {\bibfnamefont {A.}~\bibnamefont
  {Pandey}}, \bibinfo {author} {\bibfnamefont {D.~G.}\ \bibnamefont
  {Quirinale}}, \bibinfo {author} {\bibfnamefont {W.}~\bibnamefont
  {Jayasekara}}, \bibinfo {author} {\bibfnamefont {A.}~\bibnamefont {Sapkota}},
  \bibinfo {author} {\bibfnamefont {M.~G.}\ \bibnamefont {Kim}}, \bibinfo
  {author} {\bibfnamefont {R.~S.}\ \bibnamefont {Dhaka}}, \bibinfo {author}
  {\bibfnamefont {Y.}~\bibnamefont {Lee}}, \bibinfo {author} {\bibfnamefont
  {T.~W.}\ \bibnamefont {Heitmann}}, \bibinfo {author} {\bibfnamefont {P.~W.}\
  \bibnamefont {Stephens}}, \bibinfo {author} {\bibfnamefont {V.}~\bibnamefont
  {Ogloblichev}}, \bibinfo {author} {\bibfnamefont {A.}~\bibnamefont
  {Kreyssig}}, \bibinfo {author} {\bibfnamefont {R.~J.}\ \bibnamefont
  {McQueeney}}, \bibinfo {author} {\bibfnamefont {A.~I.}\ \bibnamefont
  {Goldman}}, \bibinfo {author} {\bibfnamefont {A.}~\bibnamefont {Kaminski}},
  \bibinfo {author} {\bibfnamefont {B.~N.}\ \bibnamefont {Harmon}}, \bibinfo
  {author} {\bibfnamefont {Y.}~\bibnamefont {Furukawa}}, \ and\ \bibinfo
  {author} {\bibfnamefont {D.~C.}\ \bibnamefont {Johnston}},\ }\href {\doibase
  10.1103/PhysRevB.88.014526} {\bibfield  {journal} {\bibinfo  {journal} {Phys.
  Rev. B}\ }\textbf {\bibinfo {volume} {88}},\ \bibinfo {pages} {014526}
  (\bibinfo {year} {2013})}\BibitemShut {NoStop}%
\bibitem [{\citenamefont {Li}\ \emph {et~al.}(2019{\natexlab{a}})\citenamefont
  {Li}, \citenamefont {Yin}, \citenamefont {Liu}, \citenamefont {Wang},
  \citenamefont {Xu}, \citenamefont {Song}, \citenamefont {Tian}, \citenamefont
  {Huang}, \citenamefont {Shen}, \citenamefont {Abernathy}, \citenamefont
  {Niedziela}, \citenamefont {Ewings}, \citenamefont {Perring}, \citenamefont
  {Pajerowski}, \citenamefont {Matsuda}, \citenamefont {Bourges}, \citenamefont
  {Mechthild}, \citenamefont {Su},\ and\ \citenamefont {Dai}}]{Li_2019_Dai}%
  \BibitemOpen
  \bibfield  {author} {\bibinfo {author} {\bibfnamefont {Y.}~\bibnamefont
  {Li}}, \bibinfo {author} {\bibfnamefont {Z.}~\bibnamefont {Yin}}, \bibinfo
  {author} {\bibfnamefont {Z.}~\bibnamefont {Liu}}, \bibinfo {author}
  {\bibfnamefont {W.}~\bibnamefont {Wang}}, \bibinfo {author} {\bibfnamefont
  {Z.}~\bibnamefont {Xu}}, \bibinfo {author} {\bibfnamefont {Y.}~\bibnamefont
  {Song}}, \bibinfo {author} {\bibfnamefont {L.}~\bibnamefont {Tian}}, \bibinfo
  {author} {\bibfnamefont {Y.}~\bibnamefont {Huang}}, \bibinfo {author}
  {\bibfnamefont {D.}~\bibnamefont {Shen}}, \bibinfo {author} {\bibfnamefont
  {D.~L.}\ \bibnamefont {Abernathy}}, \bibinfo {author} {\bibfnamefont {J.~L.}\
  \bibnamefont {Niedziela}}, \bibinfo {author} {\bibfnamefont {R.~A.}\
  \bibnamefont {Ewings}}, \bibinfo {author} {\bibfnamefont {T.~G.}\
  \bibnamefont {Perring}}, \bibinfo {author} {\bibfnamefont {D.~M.}\
  \bibnamefont {Pajerowski}}, \bibinfo {author} {\bibfnamefont
  {M.}~\bibnamefont {Matsuda}}, \bibinfo {author} {\bibfnamefont
  {P.}~\bibnamefont {Bourges}}, \bibinfo {author} {\bibfnamefont
  {E.}~\bibnamefont {Mechthild}}, \bibinfo {author} {\bibfnamefont
  {Y.}~\bibnamefont {Su}}, \ and\ \bibinfo {author} {\bibfnamefont
  {P.}~\bibnamefont {Dai}},\ }\href@noop {} {\bibfield  {journal} {\bibinfo
  {journal} {Phys. Rev. Lett.}\ }\textbf {\bibinfo {volume} {122}},\ \bibinfo
  {pages} {117204} (\bibinfo {year} {2019}{\natexlab{a}})}\BibitemShut
  {NoStop}%
\bibitem [{\citenamefont {Jayasekara}\ \emph {et~al.}(2013)\citenamefont
  {Jayasekara}, \citenamefont {Lee}, \citenamefont {Pandey}, \citenamefont
  {Tucker}, \citenamefont {Sapkota}, \citenamefont {Lamsal}, \citenamefont
  {Calder}, \citenamefont {Abernathy}, \citenamefont {Niedziela}, \citenamefont
  {Harmon}, \citenamefont {Kreyssig}, \citenamefont {Vaknin}, \citenamefont
  {Johnston}, \citenamefont {Goldman},\ and\ \citenamefont
  {McQueeney}}]{Jayasekara_2013}%
  \BibitemOpen
  \bibfield  {author} {\bibinfo {author} {\bibfnamefont {W.}~\bibnamefont
  {Jayasekara}}, \bibinfo {author} {\bibfnamefont {Y.}~\bibnamefont {Lee}},
  \bibinfo {author} {\bibfnamefont {A.}~\bibnamefont {Pandey}}, \bibinfo
  {author} {\bibfnamefont {G.~S.}\ \bibnamefont {Tucker}}, \bibinfo {author}
  {\bibfnamefont {A.}~\bibnamefont {Sapkota}}, \bibinfo {author} {\bibfnamefont
  {J.}~\bibnamefont {Lamsal}}, \bibinfo {author} {\bibfnamefont
  {S.}~\bibnamefont {Calder}}, \bibinfo {author} {\bibfnamefont {D.~L.}\
  \bibnamefont {Abernathy}}, \bibinfo {author} {\bibfnamefont {J.~L.}\
  \bibnamefont {Niedziela}}, \bibinfo {author} {\bibfnamefont {B.~N.}\
  \bibnamefont {Harmon}}, \bibinfo {author} {\bibfnamefont {A.}~\bibnamefont
  {Kreyssig}}, \bibinfo {author} {\bibfnamefont {D.}~\bibnamefont {Vaknin}},
  \bibinfo {author} {\bibfnamefont {D.~C.}\ \bibnamefont {Johnston}}, \bibinfo
  {author} {\bibfnamefont {A.~I.}\ \bibnamefont {Goldman}}, \ and\ \bibinfo
  {author} {\bibfnamefont {R.~J.}\ \bibnamefont {McQueeney}},\ }\href {\doibase
  10.1103/PhysRevLett.111.157001} {\bibfield  {journal} {\bibinfo  {journal}
  {Phys. Rev. Lett.}\ }\textbf {\bibinfo {volume} {111}},\ \bibinfo {pages}
  {157001} (\bibinfo {year} {2013})}\BibitemShut {NoStop}%
\bibitem [{\citenamefont {Sapkota}\ \emph {et~al.}(2017)\citenamefont
  {Sapkota}, \citenamefont {Ueland}, \citenamefont {Anand}, \citenamefont
  {Sangeetha}, \citenamefont {Abernathy}, \citenamefont {Stone}, \citenamefont
  {Niedziela}, \citenamefont {Johnston}, \citenamefont {Kreyssig},
  \citenamefont {Goldman},\ and\ \citenamefont {McQueeney}}]{Sapkota_2017}%
  \BibitemOpen
  \bibfield  {author} {\bibinfo {author} {\bibfnamefont {A.}~\bibnamefont
  {Sapkota}}, \bibinfo {author} {\bibfnamefont {B.~G.}\ \bibnamefont {Ueland}},
  \bibinfo {author} {\bibfnamefont {V.~K.}\ \bibnamefont {Anand}}, \bibinfo
  {author} {\bibfnamefont {N.~S.}\ \bibnamefont {Sangeetha}}, \bibinfo {author}
  {\bibfnamefont {D.~L.}\ \bibnamefont {Abernathy}}, \bibinfo {author}
  {\bibfnamefont {M.~B.}\ \bibnamefont {Stone}}, \bibinfo {author}
  {\bibfnamefont {J.~L.}\ \bibnamefont {Niedziela}}, \bibinfo {author}
  {\bibfnamefont {D.~C.}\ \bibnamefont {Johnston}}, \bibinfo {author}
  {\bibfnamefont {A.}~\bibnamefont {Kreyssig}}, \bibinfo {author}
  {\bibfnamefont {A.~I.}\ \bibnamefont {Goldman}}, \ and\ \bibinfo {author}
  {\bibfnamefont {R.~J.}\ \bibnamefont {McQueeney}},\ }\href@noop {} {\bibfield
   {journal} {\bibinfo  {journal} {Phys. Rev. Lett.}\ }\textbf {\bibinfo
  {volume} {119}},\ \bibinfo {pages} {147201} (\bibinfo {year}
  {2017})}\BibitemShut {NoStop}%
\bibitem [{\citenamefont {Li}\ \emph {et~al.}(2019{\natexlab{b}})\citenamefont
  {Li}, \citenamefont {Sizyuk}, \citenamefont {Sangeetha}, \citenamefont
  {Wilde}, \citenamefont {Das}, \citenamefont {Tian}, \citenamefont {Johnston},
  \citenamefont {Goldman}, \citenamefont {Kreyssig}, \citenamefont {Orth},
  \citenamefont {McQueeney},\ and\ \citenamefont {Ueland}}]{Li_2019_CaSr}%
  \BibitemOpen
  \bibfield  {author} {\bibinfo {author} {\bibfnamefont {B.}~\bibnamefont
  {Li}}, \bibinfo {author} {\bibfnamefont {Y.}~\bibnamefont {Sizyuk}}, \bibinfo
  {author} {\bibfnamefont {N.~S.}\ \bibnamefont {Sangeetha}}, \bibinfo {author}
  {\bibfnamefont {J.~M.}\ \bibnamefont {Wilde}}, \bibinfo {author}
  {\bibfnamefont {P.}~\bibnamefont {Das}}, \bibinfo {author} {\bibfnamefont
  {W.}~\bibnamefont {Tian}}, \bibinfo {author} {\bibfnamefont {D.~C.}\
  \bibnamefont {Johnston}}, \bibinfo {author} {\bibfnamefont {A.~I.}\
  \bibnamefont {Goldman}}, \bibinfo {author} {\bibfnamefont {A.}~\bibnamefont
  {Kreyssig}}, \bibinfo {author} {\bibfnamefont {P.~P.}\ \bibnamefont {Orth}},
  \bibinfo {author} {\bibfnamefont {R.~J.}\ \bibnamefont {McQueeney}}, \ and\
  \bibinfo {author} {\bibfnamefont {B.~G.}\ \bibnamefont {Ueland}},\ }\href
  {\doibase 10.1103/PhysRevB.100.024415} {\bibfield  {journal} {\bibinfo
  {journal} {Phys. Rev. B}\ }\textbf {\bibinfo {volume} {100}},\ \bibinfo
  {pages} {024415} (\bibinfo {year} {2019}{\natexlab{b}})}\BibitemShut
  {NoStop}%
\bibitem [{\citenamefont {Mao}\ and\ \citenamefont {Yin}(2018)}]{Mao_2018}%
  \BibitemOpen
  \bibfield  {author} {\bibinfo {author} {\bibfnamefont {H.}~\bibnamefont
  {Mao}}\ and\ \bibinfo {author} {\bibfnamefont {Z.}~\bibnamefont {Yin}},\
  }\href@noop {} {\bibfield  {journal} {\bibinfo  {journal} {Phys. Rev. B}\
  }\textbf {\bibinfo {volume} {98}},\ \bibinfo {pages} {115128} (\bibinfo
  {year} {2018})}\BibitemShut {NoStop}%
\bibitem [{\citenamefont {Jayasekara}\ \emph {et~al.}(2017)\citenamefont
  {Jayasekara}, \citenamefont {Pandey}, \citenamefont {Kreyssig}, \citenamefont
  {Sangeetha}, \citenamefont {Sapkota}, \citenamefont {Kothapalli},
  \citenamefont {Anand}, \citenamefont {Tian}, \citenamefont {Vaknin},
  \citenamefont {Johnston}, \citenamefont {McQueeney}, \citenamefont
  {Goldman},\ and\ \citenamefont {Ueland}}]{Jayasekara_2017}%
  \BibitemOpen
  \bibfield  {author} {\bibinfo {author} {\bibfnamefont {W.~T.}\ \bibnamefont
  {Jayasekara}}, \bibinfo {author} {\bibfnamefont {A.}~\bibnamefont {Pandey}},
  \bibinfo {author} {\bibfnamefont {A.}~\bibnamefont {Kreyssig}}, \bibinfo
  {author} {\bibfnamefont {N.~S.}\ \bibnamefont {Sangeetha}}, \bibinfo {author}
  {\bibfnamefont {A.}~\bibnamefont {Sapkota}}, \bibinfo {author} {\bibfnamefont
  {K.~K.}\ \bibnamefont {Kothapalli}}, \bibinfo {author} {\bibfnamefont
  {V.~K.}\ \bibnamefont {Anand}}, \bibinfo {author} {\bibfnamefont
  {W.}~\bibnamefont {Tian}}, \bibinfo {author} {\bibfnamefont {D.}~\bibnamefont
  {Vaknin}}, \bibinfo {author} {\bibfnamefont {D.~C.}\ \bibnamefont
  {Johnston}}, \bibinfo {author} {\bibfnamefont {R.~J.}\ \bibnamefont
  {McQueeney}}, \bibinfo {author} {\bibfnamefont {A.~I.}\ \bibnamefont
  {Goldman}}, \ and\ \bibinfo {author} {\bibfnamefont {B.~G.}\ \bibnamefont
  {Ueland}},\ }\href@noop {} {\bibfield  {journal} {\bibinfo  {journal} {Phys.
  Rev. B}\ }\textbf {\bibinfo {volume} {95}},\ \bibinfo {pages} {064425}
  (\bibinfo {year} {2017})}\BibitemShut {NoStop}%
\bibitem [{\citenamefont {Quirinale}\ \emph {et~al.}(2013)\citenamefont
  {Quirinale}, \citenamefont {Anand}, \citenamefont {Kim}, \citenamefont
  {Pandey}, \citenamefont {Huq}, \citenamefont {Stephens}, \citenamefont
  {Heitmann}, \citenamefont {Kreyssig}, \citenamefont {McQueeney},
  \citenamefont {Johnston},\ and\ \citenamefont {Goldman}}]{Quirinale_2013}%
  \BibitemOpen
  \bibfield  {author} {\bibinfo {author} {\bibfnamefont {D.~G.}\ \bibnamefont
  {Quirinale}}, \bibinfo {author} {\bibfnamefont {V.~K.}\ \bibnamefont
  {Anand}}, \bibinfo {author} {\bibfnamefont {M.~G.}\ \bibnamefont {Kim}},
  \bibinfo {author} {\bibfnamefont {A.}~\bibnamefont {Pandey}}, \bibinfo
  {author} {\bibfnamefont {A.}~\bibnamefont {Huq}}, \bibinfo {author}
  {\bibfnamefont {P.~W.}\ \bibnamefont {Stephens}}, \bibinfo {author}
  {\bibfnamefont {T.~W.}\ \bibnamefont {Heitmann}}, \bibinfo {author}
  {\bibfnamefont {A.}~\bibnamefont {Kreyssig}}, \bibinfo {author}
  {\bibfnamefont {R.~J.}\ \bibnamefont {McQueeney}}, \bibinfo {author}
  {\bibfnamefont {D.~C.}\ \bibnamefont {Johnston}}, \ and\ \bibinfo {author}
  {\bibfnamefont {A.~I.}\ \bibnamefont {Goldman}},\ }\href@noop {} {\bibfield
  {journal} {\bibinfo  {journal} {Phys. Rev. B}\ }\textbf {\bibinfo {volume}
  {88}},\ \bibinfo {pages} {174420} (\bibinfo {year} {2013})}\BibitemShut
  {NoStop}%
\bibitem [{\citenamefont {Shen}\ \emph {et~al.}(2018)\citenamefont {Shen},
  \citenamefont {Feng}, \citenamefont {Lin}, \citenamefont {Wang},\ and\
  \citenamefont {Zhong}}]{Shen_2018}%
  \BibitemOpen
  \bibfield  {author} {\bibinfo {author} {\bibfnamefont {S.}~\bibnamefont
  {Shen}}, \bibinfo {author} {\bibfnamefont {S.}~\bibnamefont {Feng}}, \bibinfo
  {author} {\bibfnamefont {Z.}~\bibnamefont {Lin}}, \bibinfo {author}
  {\bibfnamefont {Z.}~\bibnamefont {Wang}}, \ and\ \bibinfo {author}
  {\bibfnamefont {W.}~\bibnamefont {Zhong}},\ }\href@noop {} {\bibfield
  {journal} {\bibinfo  {journal} {J. Mater. Chem. C}\ }\textbf {\bibinfo
  {volume} {6}},\ \bibinfo {pages} {8076} (\bibinfo {year} {2018})}\BibitemShut
  {NoStop}%
\bibitem [{\citenamefont {Sangeetha}\ \emph {et~al.}(2019)\citenamefont
  {Sangeetha}, \citenamefont {Wang}, \citenamefont {Smirnov}, \citenamefont
  {Smetana}, \citenamefont {Mudring}, \citenamefont {Johnson}, \citenamefont
  {Tanatar}, \citenamefont {Prozorov},\ and\ \citenamefont
  {Johnston}}]{Sangeetha_2019}%
  \BibitemOpen
  \bibfield  {author} {\bibinfo {author} {\bibfnamefont {N.~S.}\ \bibnamefont
  {Sangeetha}}, \bibinfo {author} {\bibfnamefont {L.-L.}\ \bibnamefont {Wang}},
  \bibinfo {author} {\bibfnamefont {A.~V.}\ \bibnamefont {Smirnov}}, \bibinfo
  {author} {\bibfnamefont {V.}~\bibnamefont {Smetana}}, \bibinfo {author}
  {\bibfnamefont {A.-V.}\ \bibnamefont {Mudring}}, \bibinfo {author}
  {\bibfnamefont {D.~D.}\ \bibnamefont {Johnson}}, \bibinfo {author}
  {\bibfnamefont {M.~A.}\ \bibnamefont {Tanatar}}, \bibinfo {author}
  {\bibfnamefont {R.}~\bibnamefont {Prozorov}}, \ and\ \bibinfo {author}
  {\bibfnamefont {D.~C.}\ \bibnamefont {Johnston}},\ }\href@noop {} {\bibfield
  {journal} {\bibinfo  {journal} {Phys. Rev. B}\ }\textbf {\bibinfo {volume}
  {100}},\ \bibinfo {pages} {094447} (\bibinfo {year} {2019})}\BibitemShut
  {NoStop}%
\bibitem [{\citenamefont {Li}\ \emph {et~al.}(2019{\natexlab{c}})\citenamefont
  {Li}, \citenamefont {Liu}, \citenamefont {Xu}, \citenamefont {Song},
  \citenamefont {Huang}, \citenamefont {Shen}, \citenamefont {Ma},
  \citenamefont {Li}, \citenamefont {Chi}, \citenamefont {Frontzek},
  \citenamefont {Cao}, \citenamefont {Huang}, \citenamefont {Wang},
  \citenamefont {Xie}, \citenamefont {Zhang}, \citenamefont {Rong},
  \citenamefont {Shelton}, \citenamefont {Young}, \citenamefont {DiTusa},\ and\
  \citenamefont {Dai}}]{Li_2019_SrCoNi}%
  \BibitemOpen
  \bibfield  {author} {\bibinfo {author} {\bibfnamefont {Y.}~\bibnamefont
  {Li}}, \bibinfo {author} {\bibfnamefont {Z.}~\bibnamefont {Liu}}, \bibinfo
  {author} {\bibfnamefont {Z.}~\bibnamefont {Xu}}, \bibinfo {author}
  {\bibfnamefont {Y.}~\bibnamefont {Song}}, \bibinfo {author} {\bibfnamefont
  {Y.}~\bibnamefont {Huang}}, \bibinfo {author} {\bibfnamefont
  {D.}~\bibnamefont {Shen}}, \bibinfo {author} {\bibfnamefont {N.}~\bibnamefont
  {Ma}}, \bibinfo {author} {\bibfnamefont {A.}~\bibnamefont {Li}}, \bibinfo
  {author} {\bibfnamefont {S.}~\bibnamefont {Chi}}, \bibinfo {author}
  {\bibfnamefont {M.}~\bibnamefont {Frontzek}}, \bibinfo {author}
  {\bibfnamefont {H.}~\bibnamefont {Cao}}, \bibinfo {author} {\bibfnamefont
  {Q.}~\bibnamefont {Huang}}, \bibinfo {author} {\bibfnamefont
  {W.}~\bibnamefont {Wang}}, \bibinfo {author} {\bibfnamefont {Y.}~\bibnamefont
  {Xie}}, \bibinfo {author} {\bibfnamefont {R.}~\bibnamefont {Zhang}}, \bibinfo
  {author} {\bibfnamefont {Y.}~\bibnamefont {Rong}}, \bibinfo {author}
  {\bibfnamefont {W.~A.}\ \bibnamefont {Shelton}}, \bibinfo {author}
  {\bibfnamefont {D.~P.}\ \bibnamefont {Young}}, \bibinfo {author}
  {\bibfnamefont {J.~F.}\ \bibnamefont {DiTusa}}, \ and\ \bibinfo {author}
  {\bibfnamefont {P.}~\bibnamefont {Dai}},\ }\href@noop {} {\bibfield
  {journal} {\bibinfo  {journal} {Phys. Rev. B}\ }\textbf {\bibinfo {volume}
  {100}},\ \bibinfo {pages} {094446} (\bibinfo {year}
  {2019}{\natexlab{c}})}\BibitemShut {NoStop}%
\bibitem [{\citenamefont {Li}\ \emph {et~al.}(2019{\natexlab{d}})\citenamefont
  {Li}, \citenamefont {Ueland}, \citenamefont {Jayasekara}, \citenamefont
  {Abernathy}, \citenamefont {Sangeetha}, \citenamefont {Johnston},
  \citenamefont {Ding}, \citenamefont {Furukawa}, \citenamefont {Orth},
  \citenamefont {Kreyssig}, \citenamefont {Goldman},\ and\ \citenamefont
  {McQueeney}}]{Li_2019_SrINS}%
  \BibitemOpen
  \bibfield  {author} {\bibinfo {author} {\bibfnamefont {B.}~\bibnamefont
  {Li}}, \bibinfo {author} {\bibfnamefont {B.~G.}\ \bibnamefont {Ueland}},
  \bibinfo {author} {\bibfnamefont {W.~T.}\ \bibnamefont {Jayasekara}},
  \bibinfo {author} {\bibfnamefont {D.~L.}\ \bibnamefont {Abernathy}}, \bibinfo
  {author} {\bibfnamefont {N.~S.}\ \bibnamefont {Sangeetha}}, \bibinfo {author}
  {\bibfnamefont {D.~C.}\ \bibnamefont {Johnston}}, \bibinfo {author}
  {\bibfnamefont {Q.-P.}\ \bibnamefont {Ding}}, \bibinfo {author}
  {\bibfnamefont {Y.}~\bibnamefont {Furukawa}}, \bibinfo {author}
  {\bibfnamefont {P.~P.}\ \bibnamefont {Orth}}, \bibinfo {author}
  {\bibfnamefont {A.}~\bibnamefont {Kreyssig}}, \bibinfo {author}
  {\bibfnamefont {A.~I.}\ \bibnamefont {Goldman}}, \ and\ \bibinfo {author}
  {\bibfnamefont {R.~J.}\ \bibnamefont {McQueeney}},\ }\href {\doibase
  10.1103/PhysRevB.100.054411} {\bibfield  {journal} {\bibinfo  {journal}
  {Phys. Rev. B}\ }\textbf {\bibinfo {volume} {100}},\ \bibinfo {pages}
  {054411} (\bibinfo {year} {2019}{\natexlab{d}})}\BibitemShut {NoStop}%
\bibitem [{\citenamefont {Kreyssig}\ \emph {et~al.}(2007)\citenamefont
  {Kreyssig}, \citenamefont {Chang}, \citenamefont {Janssen}, \citenamefont
  {Kim}, \citenamefont {Nandi}, \citenamefont {Yan}, \citenamefont {Tan},
  \citenamefont {McQueeney}, \citenamefont {Canfield},\ and\ \citenamefont
  {Goldman}}]{Kreyssig_2007}%
  \BibitemOpen
  \bibfield  {author} {\bibinfo {author} {\bibfnamefont {A.}~\bibnamefont
  {Kreyssig}}, \bibinfo {author} {\bibfnamefont {S.}~\bibnamefont {Chang}},
  \bibinfo {author} {\bibfnamefont {Y.}~\bibnamefont {Janssen}}, \bibinfo
  {author} {\bibfnamefont {J.~W.}\ \bibnamefont {Kim}}, \bibinfo {author}
  {\bibfnamefont {S.}~\bibnamefont {Nandi}}, \bibinfo {author} {\bibfnamefont
  {J.~Q.}\ \bibnamefont {Yan}}, \bibinfo {author} {\bibfnamefont
  {L.}~\bibnamefont {Tan}}, \bibinfo {author} {\bibfnamefont {R.~J.}\
  \bibnamefont {McQueeney}}, \bibinfo {author} {\bibfnamefont {P.~C.}\
  \bibnamefont {Canfield}}, \ and\ \bibinfo {author} {\bibfnamefont {A.~I.}\
  \bibnamefont {Goldman}},\ }\href@noop {} {\bibfield  {journal} {\bibinfo
  {journal} {Phys. Rev. B}\ }\textbf {\bibinfo {volume} {76}},\ \bibinfo
  {pages} {054421} (\bibinfo {year} {2007})}\BibitemShut {NoStop}%
\bibitem [{\citenamefont {Sangeetha}\ \emph {et~al.}(2017)\citenamefont
  {Sangeetha}, \citenamefont {Smetana}, \citenamefont {Mudring},\ and\
  \citenamefont {Johnston}}]{Sangeetha_2017}%
  \BibitemOpen
  \bibfield  {author} {\bibinfo {author} {\bibfnamefont {N.~S.}\ \bibnamefont
  {Sangeetha}}, \bibinfo {author} {\bibfnamefont {V.}~\bibnamefont {Smetana}},
  \bibinfo {author} {\bibfnamefont {A.~V.}\ \bibnamefont {Mudring}}, \ and\
  \bibinfo {author} {\bibfnamefont {D.~C.}\ \bibnamefont {Johnston}},\
  }\href@noop {} {\bibfield  {journal} {\bibinfo  {journal} {Phys. Rev. Lett.}\
  }\textbf {\bibinfo {volume} {119}},\ \bibinfo {pages} {257203} (\bibinfo
  {year} {2017})}\BibitemShut {NoStop}%
\bibitem [{\citenamefont {Hoffman}\ and\ \citenamefont
  {Zheng}(1985)}]{Hoffman_1985}%
  \BibitemOpen
  \bibfield  {author} {\bibinfo {author} {\bibfnamefont {R.}~\bibnamefont
  {Hoffman}}\ and\ \bibinfo {author} {\bibfnamefont {C.}~\bibnamefont
  {Zheng}},\ }\href@noop {} {\bibfield  {journal} {\bibinfo  {journal} {J.
  Phys. Chem.}\ }\textbf {\bibinfo {volume} {89}},\ \bibinfo {pages} {4175}
  (\bibinfo {year} {1985})}\BibitemShut {NoStop}%
\bibitem [{\citenamefont {Johnston}(2012)}]{Johnston_2012}%
  \BibitemOpen
  \bibfield  {author} {\bibinfo {author} {\bibfnamefont {D.~C.}\ \bibnamefont
  {Johnston}},\ }\href@noop {} {\bibfield  {journal} {\bibinfo  {journal}
  {Phys. Rev. Lett.}\ }\textbf {\bibinfo {volume} {109}},\ \bibinfo {pages}
  {077201} (\bibinfo {year} {2012})}\BibitemShut {NoStop}%
\bibitem [{\citenamefont {Johnston}(2015)}]{Johnston_2015}%
  \BibitemOpen
  \bibfield  {author} {\bibinfo {author} {\bibfnamefont {D.~C.}\ \bibnamefont
  {Johnston}},\ }\href@noop {} {\bibfield  {journal} {\bibinfo  {journal}
  {Phys. Rev. B}\ }\textbf {\bibinfo {volume} {91}},\ \bibinfo {pages} {064427}
  (\bibinfo {year} {2015})}\BibitemShut {NoStop}%
\bibitem [{\citenamefont {Bak}\ and\ \citenamefont {von
  Boehm}(1980)}]{Bak_1980}%
  \BibitemOpen
  \bibfield  {author} {\bibinfo {author} {\bibfnamefont {P.}~\bibnamefont
  {Bak}}\ and\ \bibinfo {author} {\bibfnamefont {J.}~\bibnamefont {von
  Boehm}},\ }\href {\doibase 10.1103/PhysRevB.21.5297} {\bibfield  {journal}
  {\bibinfo  {journal} {Phys. Rev. B}\ }\textbf {\bibinfo {volume} {21}},\
  \bibinfo {pages} {5297} (\bibinfo {year} {1980})}\BibitemShut {NoStop}%
\bibitem [{\citenamefont {Fisher}\ and\ \citenamefont
  {Selke}(1980)}]{Fisher_1980}%
  \BibitemOpen
  \bibfield  {author} {\bibinfo {author} {\bibfnamefont {M.~E.}\ \bibnamefont
  {Fisher}}\ and\ \bibinfo {author} {\bibfnamefont {W.}~\bibnamefont {Selke}},\
  }\href {\doibase 10.1103/PhysRevLett.44.1502} {\bibfield  {journal} {\bibinfo
   {journal} {Phys. Rev. Lett.}\ }\textbf {\bibinfo {volume} {44}},\ \bibinfo
  {pages} {1502} (\bibinfo {year} {1980})}\BibitemShut {NoStop}%
\bibitem [{\citenamefont {Villain}\ and\ \citenamefont
  {Gordon}(1980)}]{Villain_1980}%
  \BibitemOpen
  \bibfield  {author} {\bibinfo {author} {\bibfnamefont {J.}~\bibnamefont
  {Villain}}\ and\ \bibinfo {author} {\bibfnamefont {M.~B.}\ \bibnamefont
  {Gordon}},\ }\href {\doibase 10.1088/0022-3719/13/17/005} {\bibfield
  {journal} {\bibinfo  {journal} {J. Phys. C: Solid State Phys.}\ }\textbf
  {\bibinfo {volume} {13}},\ \bibinfo {pages} {3117} (\bibinfo {year}
  {1980})}\BibitemShut {NoStop}%
\end{thebibliography}%

 \end{document}